\documentclass[12pt]{article}

\pdfoutput=1

\usepackage{graphicx}

\usepackage{hyperref}
\usepackage{amsmath}
\usepackage[table]{xcolor}
\usepackage{color}
\begin{document}

\hoffset = -0.3truecm
\voffset = -1.1truecm

\title{\bf
Electrically Charged One and a Half Monopole Solution\footnote{To be submitted for publication}}

\author{
{\bf Rosy Teh\footnote{E-mail: rosyteh@usm.my}}, 
{\bf Ban-Loong Ng and Khai-Ming Wong}\\
{\normalsize School of Physics, Universiti Sains Malaysia}\\
{\normalsize 11800 USM Penang, Malaysia}}

\date{April 28, 2014}
\maketitle

\begin{abstract}
Recently, we have discussed the coexistence of a finite energy one-half monopole and a 't Hooft-Polyakov monopole of opposite magnetic charges. In this paper, we would like to introduce electric charge into this new monopoles configuration, thus creating a one and a half dyon. This new dyon possesses finite energy, magnetic dipole moment and angular momentum and is able to precess in the presence of an external magnetic field. Similar to the other dyon solutions, when the Higgs self-coupling constant, $\lambda$, is nonvanishing, this new dyon solution possesses critical electric charge, total energy, magnetic dipole moment, and dipole separation as the electric charge parameter, $\eta$, approaches one. The electric charge and total energy increase with $\eta$ to maximum critical values as $\eta\rightarrow1$ for all nonvanishing $\lambda$. However, the magnetic dipole moment decreases with $\eta$ when $\lambda\geq0.1$ and the dipole separation decreases with $\eta$ when $\lambda\geq1$ to minimum critical values as $\eta\rightarrow1$. 
\end{abstract}


\section{Introduction}

The SU(2) Yang-Mills-Higgs (YMH) field theory possesses a rich spectrum of monopoles configurations which are invariant under a U(1) subgroup of the local SU(2) gauge group. The invariance of the U(1) subgroup is an important part of the theory as upon symmetry breaking it will give rise to Maxwell's electromagnetic field theory \cite{kn:1}. Some of the well known work on monopole solutions are listed in Ref. \cite{kn:1} to \cite{kn:3}. The 't Hooft-Polyakov monopole solution is a numerical solution \cite{kn:1}-\cite{kn:2}, whereas exact solutions can be obtained only in the Bogomol'nyi-Prasad-Sommerfield (BPS) limit when the Higgs potential vanishes \cite{kn:3}. Numerical BPS monopole solutions were discussed in Ref. \cite{kn:4}, whereas numerical solutions with axial symmetry and nonvanishing Higgs potential were given in Ref. \cite{kn:5}. Recently, numerical generalized Jacobi elliptic single monopole solutions and numerical generalized Jacobi elliptic MAP and single vortex ring solutions were also discussed in Ref. \cite{kn:6}. The monopole solutions discussed in Ref. \cite{kn:1} to \cite{kn:6} possessed integer topological magnetic charge. 

However there are also papers with discussions on particles with one-half monopole magnetic charge. These include the work of Harikumar et al. \cite{kn:7} where they demonstrated the existence of generic smooth Yang-Mills (YM) potentials of one-half monopoles. Exact axially symmetric and mirror symmetric one-half monopole solutions with Dirac-like string were discussed in Ref. \cite{kn:8}. However these exact solutions possess infinite total energy. 

Recently, axially symmetric, finite energy particles of one-half monopole magnetic charge \cite{kn:9} and particles of positive one and negative half monopole magnetic charges \cite{kn:10} were shown to exist. The 't Hooft magnetic fields of these solutions at spatial infinity correspond to the magnetic field of a positive one-half magnetic monopole located at the origin, $r=0$, and a semi-infinite Dirac string located on one half of the $z$-axis which carries magnetic flux of $\frac{2\pi}{g}$ from infinity to the origin, thus making the net magnetic charge of the configuration zero. The non-Abelian solutions possess gauge potentials that are singular only along one-half of the $z$-axis, elsewhere they are regular \cite{kn:11}. The total energies of these new magnetic monopole solutions were found to increase with $\lambda$. 

A dyon is  a particle that possesses both magnetic and electric charges. A dyon with a fixed magnetic charge can possess varying electric charges \cite{kn:12} at the classical level. The dyons solutions of Julia and Zee \cite{kn:13} are time independent solutions that possess nonvanishing kinetic energy. The Julia-Zee solutions are non-self-dual even in the BPS limit when the electric charge is nonvanishing. The exact dyon solutions found by Prasad and Sommerfield \cite{kn:13} are actually Julia and Zee dyon solutions in the BPS limit. These solutions are stable as they are the absolute minima of the energy \cite{kn:14}.

All the monopole solutions of the SU(2) YMH theory can acquire an electric charge to become a dyon as shown by the work of Ref. \cite{kn:13}. Axially symmetrical single pole dyons were constructed by B. Hartmann et al. \cite{kn:15}. These axial dyons are actually generalized 't Hooft-Polyakov monopoles that possess magnetic charges, $n=1, 2, 3$ and nonvanishing electric charges. It was found in Ref. \cite{kn:15} that when the strength of the Higgs potential $\lambda$ is nonvanishing, the total electric charge and total energy of the system approach finite critical (maximum) values when the electric charge parameter, $\eta$, approaches one. However when $\lambda=0$, the total electric charge and total energy approach infinity when the parameter $\eta$ approaches one. Similarly when $\lambda$ is nonvanishing, the electric charge, total energy, and magnetic dipole moment of the zero topological charge sector of the monopole-antimonopole pair (MAP) and vortex ring solutions  \cite{kn:16} and the one-half dyon solution  \cite{kn:17} approach finite critical (maximum) values when the electric charge parameter $\eta$ approaches one. However when $\lambda=0$, the electric charge, total energy, and magnetic dipole moment approach infinity when $\eta$ approaches one. The MAP dyons were also investigated in Ref. \cite{kn:18} by varying $\lambda$ for fixed value of $\eta$.

Since the one and a half monopoles solution of Ref. \cite{kn:10} is a new finite energy solution with properties that differ from the usual monopoles, we would like to further study its properties and behaviour when electric charges are introduced into the configuration. This is done by using the standard procedure of Julia and Zee \cite{kn:13} for the magnetic ansatz \cite{kn:15} - \cite{kn:18}. We calculate numerically for the dimensionless electric charge $Q$, total energy $E$, angular momentum $J_z$ about the $z$-axis of symmetry, magnetic dipole moment $\mu_m$, and dipole separation, $d_z$, of the new dyon solution when the electric charge parameter $\eta$ is varied from zero to one and when the Higgs self-coupling constant is varied from zero to 12. Since this new dyon possesses finite magnetic dipole moment and angular momentum, it is able to precess in the presence of an external magnetic field. 

Similar to the single pole dyon \cite {kn:15} and the MAP dyons \cite{kn:16}, \cite{kn:18}, this new dyon possesses critical (maximum) electric charge and total energy when the Higgs self-coupling constant, $\lambda$, is nonvanishing  and the electric charge parameter, $\eta$, approaches one. When $\lambda$ vanishes, these quantities approach infinity when $\eta$ approaches one. However in contrast to the other previous dyon solutions, the magnetic dipole moment decreases with $\eta$ when $\lambda\geq0.1$ and the dipole separation decreases with $\eta$ when $\lambda\geq1$ to minimum critical values as $\eta\rightarrow1$. 

We also calculate for the total energy $E$, the total electric charge $Q$, the magnetic dipole moment $\mu_m$, and dipole separation, $d_z$, of this new dyon solution for fixed values of $\eta$ and for $\lambda$ from zero to 12. In general, the total electric charge, magnetic dipole moment, and dipole separation decrease exponentially with increasing $\lambda^{1/2}$. The total energy for small values of $\eta< 0.7$ however increases logarithmically with increasing $\lambda^{1/2}$. The total energy for larger values of $0.7 <\eta\leq 1$ first decreases for small values of $\lambda^{1/2}$ then increases with $\lambda^{1/2}$.

We briefly review the SU(2) Yang-Mills-Higgs field theory in the next section. In section 3, we discussed on the construction of the new dyon solution. The magnetic ansatz used in obtaining the new dyon solution and some of its basic properties are given in this section.  The numerical results of our calculations of the new dyon solution are presented and discussed in section 4. We end with some comments in section 5.

\section{The SU(2) Yang-Mills-Higgs Theory}
\label{sect:2}
The Lagrangian in this 3+1 dimensionional theory is
\begin{equation}
{\cal L} = -\frac{1}{4}F^a_{\mu\nu} F^{a\mu\nu} - \frac{1}{2}D^\mu \Phi^a D_\mu \Phi^a - \frac{1}{4}\lambda(\Phi^a\Phi^a - \xi^2)^2. 
\label{eq.1}
\end{equation}

\noindent where the first two terms on the left-hand side Eq. (\ref{eq.1}) are the kinetic energy terms and the last term is the nonvanishing Higgs potential.
 Here the Higgs field mass is $\mu$ and the strength of the Higgs potential is $\lambda$ which are constants. The vacuum expectation value of the Higgs field is $\xi=\mu/\sqrt{\lambda}$. 
The covariant derivative of the Higgs field and the gauge field strength tensor are given respectively by 

\begin{eqnarray}
D_{\mu}\Phi^{a} &=& \partial_{\mu} \Phi^{a} + g\epsilon^{abc} A^{b}_{\mu}\Phi^{c},\nonumber\\
F^a_{\mu\nu} &=& \partial_{\mu}A^a_\nu - \partial_{\nu}A^a_\mu + g\epsilon^{abc}A^b_{\mu}A^c_\nu,
\label{eq.2}
\end{eqnarray}
where $g$ is the gauge field coupling constant. The metric used is $g_{\mu\nu} = (- + + +)$. The SU(2) internal group indices $a, b, c = 1, 2, 3$  and the space-time indices are $\mu, \nu, \alpha = 0, 1, 2$, and $3$ in Minkowski space. The equations of motion that follow from the Lagrangian (\ref{eq.1}) are
\begin{eqnarray}
D^{\mu}F^a_{\mu\nu} &=& \partial^{\mu}F^a_{\mu\nu} + g\epsilon^{abc}A^{b\mu}F^c_{\mu\nu} = g\epsilon^{abc}\Phi^{b}D_{\nu}\Phi^c,\nonumber\\
D^{\mu}D_{\mu}\Phi^a &=& \lambda\Phi^a(\Phi^{b}\Phi^{b} - \xi^2).
\label{eq.3}
\end{eqnarray}
In the limit of vanishing $\mu$ and $\lambda$, the Higgs potential vanishes and self-dual solutions can be obtained by solving the first order partial differential Bogomol'nyi equation, 
\begin{equation}
B^a_i \pm D_i \Phi^a = 0, ~~~\mbox{where}~ B^a_i=-\frac{1}{2}\epsilon_{ijk}F^a_{jk}.
\label{eq.4}
\end{equation}

The electromagnetic field tensor proposed by 't Hooft \cite{kn:2} upon symmetry breaking is
\begin{eqnarray}
F_{\mu\nu} &=& \hat{\Phi}^a F^a_{\mu\nu} - \frac{1}{g}\epsilon^{abc}\hat{\Phi}^{a}D_{\mu}\hat{\Phi}^{b}D_{\nu}\hat{\Phi}^c,\nonumber\\
	&=& \partial_{\mu}A_\nu - \partial_{\nu}A_\mu - \frac{1}{g}\epsilon^{abc}\hat{\Phi}^{a}\partial_{\mu}\hat{\Phi}^{b}\partial_{\nu}\hat{\Phi}^c = G_{\mu\nu}+H_{\mu\nu}, ~~~\mbox{where}
\label{eq.5}\\
G_{\mu\nu} &=&  \partial_{\mu}A_\nu - \partial_{\nu}A_\mu, ~\mbox{and}~ H_{\mu\nu} = - \frac{1}{g}\epsilon^{abc}\hat{\Phi}^{a}\partial_{\mu}\hat{\Phi}^{b}\partial_{\nu}\hat{\Phi}^c,
\label{eq.6}
\end{eqnarray}

\noindent are the non-topological Maxwell part and the topological Dirac part of the electromagnetic field respectively. Here $A_\mu = \hat{\Phi}^{a}A^a_\mu$, the Higgs unit vector, $\hat{\Phi}^a = \Phi^a/|\Phi|$, and the Higgs field magnitude $|\Phi| = \sqrt{\Phi^{a}\Phi^{a}}$. This 't Hooft electromagnetic field is precisely the restricted field strength that we obtain from the gauge independent Abelian projection known as the Cho projection \cite{kn:19}
Hence the decomposed magnetic field is
\begin{eqnarray}
B_i = -\frac{1}{2}\epsilon_{ijk}F_{jk}
      = B^G_i + B^H_i,
\label{eq.7}
\end{eqnarray}
where $B_i^G$ and $B_i^H$ are the gauge part and Higgs part of the magnetic field respectively. The net magnetic charge of the system is
\begin{eqnarray}
M = \frac{1}{4\pi} \int \partial^i B_i ~d^{3}x  = \frac{1}{4\pi} \oint d^{2}\sigma_{i}~B_i.
\label{eq.8}
\end{eqnarray}

Since the topological magnetic current is \cite{kn:20} 
$k_\mu = \frac{1}{8\pi}~\epsilon_{\mu\nu\rho\sigma}~\epsilon_{abc}~\partial^{\nu}\hat{\Phi}^{a}~\partial^{\rho}\hat{\Phi}^{b}~\partial^{\sigma}\hat{\Phi}^c$,
and 
\begin{eqnarray}
M_H & = & \frac{1}{g}\int d^{3}x~k_0 = \frac{1}{4\pi} \oint d^{2}\sigma_{i}~B_i^H,
\label{eq.9}
\end{eqnarray}

\noindent hence $M_H$ is the corresponding conserved topological magnetic charge.
The magnetic charge $M_H$ is the total magnetic charge of the system if and only if the gauge field is nonsingular \cite{kn:21}. If the gauge field is singular and carries Dirac string monopoles, then the magnetic charge carried by the gauge field is

\begin{eqnarray}
M_G & = & -\frac{1}{8\pi}\oint d^{2}\sigma_{i}\epsilon_{ijk}\left(\partial_j A_k - \partial_k A_j\right)\nonumber\\
& = & \frac{1}{4\pi} \oint d^{2}\sigma_{i}~B_i^G, 
\label{eq.10}
\end{eqnarray}

\noindent and the total magnetic charge of the system is $M = M_G + M_H$.


\section{The Magnetic Ansatz}
\label{sect:3}

The magnetic ansatz used are given by, \cite{kn:17}
\begin{eqnarray}
gA_i^a &=&  - \frac{1}{r}\psi_1(r, \theta) \hat{n}^{a}_\phi\hat{\theta}_i + \frac{1}{r\sin\theta}P_1(r, \theta)\hat{n}^{a}_\theta\hat{\phi}_i
+ \frac{1}{r}R_1(r, \theta)\hat{n}^{a}_\phi\hat{r}_i - \frac{1}{r\sin\theta}P_2(r, \theta)\hat{n}^{a}_r\hat{\phi}_i, \nonumber\\
gA^a_0 &=& \tau_1(r, \theta)~\hat{n}^a_r + \tau_2(r, \theta)\hat{n}^a_\theta,,  \nonumber\\
g\Phi^a &=& \Phi_1(r, \theta)~\hat{n}^a_r + \Phi_2(r, \theta)\hat{n}^a_\theta,
\label{eq.11}
\end{eqnarray}
\noindent where $P_1(r, \theta)=\sin\theta~\psi_2(r, \theta)$ and $P_2(r, \theta)=\sin\theta~R_2(r, \theta)$. 
The spatial spherical coordinate orthonormal unit vectors are
\begin{eqnarray}
\hat{r}_i &=& \sin\theta ~\cos \phi ~\delta_{i1} + \sin\theta ~\sin \phi ~\delta_{i2} + \cos\theta~\delta_{i3}, \nonumber\\
\hat{\theta}_i &=& \cos\theta ~\cos \phi ~\delta_{i1} + \cos\theta ~\sin \phi ~\delta_{i2} - \sin\theta ~\delta_{i3}, \nonumber\\
\hat{\phi}_i &=& -\sin \phi ~\delta_{i1} + \cos \phi ~\delta_{i2},
\label{eq.12}
\end{eqnarray}
and the isospin coordinate orthonormal unit vectors are 
\begin{eqnarray}
\hat{n}_r^a &=& \sin \theta ~\cos n\phi ~\delta_{1}^a + \sin \theta ~\sin n\phi ~\delta_{2}^a + \cos \theta~\delta_{3}^a,\nonumber\\
\hat{n}_\theta^a &=& \cos \theta ~\cos n\phi ~\delta_{1}^a + \cos \theta ~\sin n\phi ~\delta_{2}^a - \sin \theta ~\delta_{3}^a,\nonumber\\
\hat{n}_\phi^a &=& -\sin n\phi ~\delta_{1}^a + \cos n\phi ~\delta_{2}^a; ~~~\mbox{where}~~n\geq 1.
\label{eq.13}
\end{eqnarray}
The $\phi$-winding number $n$ is in general a natural number. However in our work here, we take $n=1$.



The general Higgs fields in the spherical and the rectangular coordinate systems are
\begin{eqnarray}
g\Phi^a &=& \Phi_1(x)~\hat{n}^a_r + \Phi_2(x)\hat{n}^a_\theta + \Phi_3(x)\hat{n}^a_\phi\nonumber\\
&=& \tilde{\Phi}_1(x) ~\delta^{a1} + \tilde{\Phi}_2(x) ~\delta^{a2} + \tilde{\Phi}_3(x) ~\delta^{a3},
\label{eq.14}
\end{eqnarray}
respectively, where
\begin{eqnarray}
\tilde{\Phi}_1 &=& \sin\theta \cos n\phi ~\Phi_1 + \cos\theta \cos n\phi ~\Phi_2 - \sin n\phi ~\Phi_3
= |\Phi|\sin\alpha \cos\beta\nonumber\\
\tilde{\Phi}_2 &=& \sin\theta \sin n\phi ~\Phi_1 + \cos\theta \sin n\phi ~\Phi_2 + \cos n\phi ~\Phi_3
= |\Phi|\sin\alpha \sin\beta\nonumber\\
\tilde{\Phi}_3 &=& \cos\theta ~\Phi_1 - \sin\theta ~\Phi_2 = |\Phi|\cos\alpha.
\label{eq.15}
\end{eqnarray}
The axially symmetric Higgs unit vector in the rectangular coordinate system is
\begin{eqnarray}
\hat{\Phi}^a &=& \sin\alpha \cos\beta ~\delta^{a1} + \sin\alpha \sin\beta ~\delta^{a2} + \cos\alpha ~\delta^{a3}, 
\label{eq.16}\\
\cos\alpha &=& h_1\cos\theta - h_2\sin\theta,~~~\sin\alpha = h_1\sin\theta + h_2\cos\theta,\nonumber\\
h_1 &=& \frac{\Phi_1}{|\Phi|}, ~~~h_2 = \frac{\Phi_2}{|\Phi|},~~~ \beta=n\phi.
\label{eq.17}
\end{eqnarray}



By using the definition of $\cos\alpha$ (\ref{eq.17}), the Higgs part of the 't Hooft magnetic field (\ref{eq.7}) can be reduced to
\begin{eqnarray}
gB_i^H = -n\epsilon_{ijk} \partial^j\cos\alpha\partial^k\phi.
\label{eq.18}
\end{eqnarray}
The gauge part of the magnetic field (\ref{eq.7}) can be written in similar form
\begin{eqnarray}
gB^G_i = -n\epsilon_{ijk}\partial_j\cos\kappa ~\partial_k \phi, ~~~\cos\kappa = \frac{1}{n}\left(h_2P_1 - h_1P_2\right).
 \label{eq.19}
\end{eqnarray}

\noindent Hence the 't Hooft's magnetic field which is the sum of the Higgs part (\ref{eq.18}) and the gauge part (\ref{eq.19}) is given by
\begin{eqnarray}
gB_i = -n\epsilon_{ijk}\partial_j(\cos\alpha + \cos\kappa)~\partial_k \phi = -\epsilon_{ijk}\partial_j{\cal A}_k,
\label{eq.20}
\end{eqnarray}
where ${\cal A}_i$ is the 't Hooft's gauge potential.
The magnetic field lines of the configuration can be plotted by drawing the contour lines of $(\cos\alpha + \cos\kappa) = $ constant on the vertical plane $\phi=0$ as shown in Figure \ref{fig.1} (c). The orientation of the magnetic field can also be plotted by using the vector field plot of the magnetic field unit vector as shown in Figure \ref{fig.1} (d),
\begin{eqnarray}
\hat{B}_i = \frac{-\partial_\theta (\cos\alpha + \cos\kappa)\hat{r}_i + r \partial_r (\cos\alpha + \cos\kappa)\hat{\theta}_i}{\sqrt{[r \partial_r (\cos\alpha + \cos\kappa)]^2 + [\partial_\theta (\cos\alpha + \cos\kappa)]^2}}.
\label{eq.21}
\end{eqnarray}

At spatial infinity in the Higgs vacuum, all the non-Abelian components of the gauge potential vanish and the non-Abelian electromagnetic field tends to 
\begin{eqnarray}
\left.F^a_{\mu\nu}\right|_{r\rightarrow\infty} &=& \{\partial_\mu A_\nu - \partial_\nu A_\mu - \frac{1}{g}\epsilon^{cde}\hat{\Phi}^c\partial_\mu \hat{\Phi}^d \partial_\nu \hat{\Phi}^e\}\hat{\Phi}^a \nonumber\\
						&=& F_{\mu\nu}\hat{\Phi}^a, 
\label{eq.22}
\end{eqnarray}
where $F_{\mu\nu}$ is just the 't Hooft electromagnetic field. However there is no unique way of representing the Abelian electromagnetic field in the region of the monopole outside the Higgs vacuum at finite values of $r$ \cite{kn:22}. One proposal was given by 't Hooft as in Eq. (\ref{eq.5}) and 
another was given by Bogomol'nyi \cite{kn:2} and Faddeev \cite{kn:23}. In the latter definition which is less singular, the magnetic and electric fields are given respectively by
\begin{eqnarray}
{\cal B}_i = B_i^a \left(\frac{\Phi^a}{\xi}\right), ~\mbox{and}~ ~{\cal E}_i = E_i^a \left(\frac{\Phi^a}{\xi}\right),
\label{eq.23}
\end{eqnarray}
where $\xi$ is the vacuum expectation value of the Higgs field. 
With this definition of the electromagnetic field (\ref{eq.23}), there will be a magnetic charge density distribution contributed by the non-Abelian components of the gauge field in the finite $r$ region. Since the magnetic charge density of the one-half dyon solution is singular and yet integrable, we therefore define the weighted magnetic charge density to be ${\cal M} = \frac{1}{2}r^2\sin\theta\{\partial^i {\cal B}_i\}$ which can be plotted as in Figure \ref{fig.2} (a). 
In the Higgs vacuum at spatial infinity, both definitions of the electromagnetic field (\ref{eq.5}) and (\ref{eq.23}) become similar.

We can also evaluate numerically the different magnetic charges at different distances $r$ from the origin by the following definitions,
\begin{eqnarray}
&&M_{\{UH\}} = -\frac{1}{2g}\left.\left\{ \cos \alpha + \cos\kappa \right\} \right|^{\theta=\frac{1}{2}\pi}_{\theta=0, r}, ~~
M_{\{LH\}} = -\frac{1}{2g}\left.\left\{ \cos \alpha + \cos\kappa \right\} \right|^{\theta=\pi}_{\theta=\frac{1}{2}\pi, r},  \nonumber\\
&&M_G = -\frac{1}{2g}\left.\left\{\cos\kappa \right\}\right|^{\theta=\pi}_{\theta=0, r}, ~~
M_H = -\frac{1}{2g}\left.\left\{ \cos \alpha\right\}\right|^{\theta=\pi}_{\theta=0, r}, ~~M=M_G + M_H.
\label{eq.24}
\end{eqnarray}
where $M_{\{UH\}}$ and $M_{\{LH\}}$ are the magnetic charges covered by the upper and lower hemispheres respectively at distances $r$ from the origin. 



Using the definition (\ref{eq.23}), we similarly define the weighted electric charge density to be ${\cal Q} =\frac{1}{2}r^2\sin\theta\{\partial^i {\cal E}_i\}$. The weighted electric charge density distribution ${\cal Q}$ can be calculated and plotted numerically as in Figure \ref{fig.2} (b), where ${\cal E}_i = {\cal F}_{i0}$ is the electric field.
The weighted electric charge density, ${\cal Q}$, of the dyon solutions is solely positive throughout space when $\eta$ is positive.  

At spatial infinity in the Higgs vacuum, 
\begin{eqnarray}
{\cal E}_i = E_i = F_{i0}= \partial_i A_0 = \partial_i\left\{\tau_1 \cos(\alpha-\theta) + \tau_2 ~\sin(\alpha-\theta)\right\} = \partial_i |\tau|,
\label{eq.25}                    
\end{eqnarray}
where $|\tau|=\sqrt{\tau_1^2+\tau_2^2}$, since the time component of the gauge field, $A^a_0$, is assumed parallel to the Higgs field, $\Phi^a$, in isospin space \cite{kn:15} - \cite{kn:17}.
Hence both definitions for the electromagnetic field strength will give the same total electric charge, $Q(\lambda, \eta)$, with ${\cal E}_i $ less singular than $E_i$ at finite values of $r$.
Unlike the magnetic field, the electric field varies proportionally with the constant $0\leq\eta< 1$. The electric field can therefore be switched off by setting $\eta=0$. 
The contour plot of the time component of the gauge potential, $A_0=$ constant, shown in Figure \ref{fig.1} (a) shows the line of equipotential of the electric field.
The 2D vector field plot of the electric field unit vector, $\hat{E}_i = \partial_i A_0/\sqrt{\partial_i A_0 \partial_i A_0}$ is shown in Figure \ref{fig.1} (b).

From Gauss' law, the total electric charge of the dyon in unit of $4\pi\xi$ is given by,
$Q(\lambda, \eta) = \frac{1}{4\pi\xi}\int_{r\rightarrow \infty}{\cal E}_i\hat{r}_i~r^2\sin\theta~d\theta d\phi$.
Since we assume that, $A^a_0$, is parallel to the Higgs field at large $r$, then ,
$Q(\lambda, \eta) =\frac{1}{\xi} \lim_{r\rightarrow \infty} r^2\partial_r |\tau|$,
can be calculated numerically.
An alternative way to find $Q$ is to assume that, $|A_0^a|= |\tau|\rightarrow \eta\xi(1-\frac{a_1}{r})$, where $a_1$ is a constant, at large $r$. Then $Q=\eta \xi a_1$ can be obtained by plotting $r(|\tau|-\eta\xi)$ and reading off the value of $\eta\xi a_1$ at large $r$. In our case, we choose to evaluate $Q$ by numerically evaluating the volume integration  
\begin{eqnarray}
Q=\frac{1}{4\pi\xi}\int{\partial^i{\cal E}_i}~d^3x.
\label{eq.26}                    
\end{eqnarray}



From Maxwell electromagnetic theory, the 't Hooft's gauge potential, ${\cal A}_i$, of Eq. (\ref{eq.20}) at large $r$ tends to
\begin{eqnarray}
{\cal A}_i &=& (\cos\alpha + \cos\kappa)\partial_i\phi |_{r\rightarrow\infty} = \frac{\hat{\phi}_i}{r\sin\theta}\left\{\frac{1}{2}(\cos\theta+1) + \frac{F_G(\theta)}{r}\right\},
\label{eq.27}\\
F_G(\theta) &=&-\mu_m\sin^2\theta + \nu_m\left\{\sin^2\theta \ln\left|\frac{1+\cos\theta}{\sin\theta}\right| + \cos\theta\right\},
\label{eq.28}
\end{eqnarray}
where $\mu_m$ is the dimensionless magnetic dipole moment of the one-half monopole.
From the numerical solution, $F_G(\theta)$ can be calculated numerically using the expression,
\begin{eqnarray}
F_G(\theta) = r\{h_2(P_1-\sin\theta) - h_1(P_2-\cos\theta) - \frac{1}{2}(\cos\theta + 1)\}|_{r\rightarrow\infty}.
\label{eq.29}
\end{eqnarray}
Plotting the graphs of $F_G(\theta)$ versus angle $\theta$, we find that $F_G(\theta) =-\mu_m\sin^2\theta$, and $\mu_m$ is nonvanishing for all values of $\lambda$ and $0\leq \eta< 1$. The constant, $\nu_m$, is however zero. The value of $\mu_m$ is read from the graphs of $F_G(\theta)$ versus angle $\theta$ at $\theta=\frac{\pi}{2}$. The magnetic dipole moment, $\mu_m$, were obtained for various values of $0\leq \eta< 1$ and $0<\lambda\leq 12$ (Table \ref{table.1} and \ref{table.2}).

From the energy momentum tensor of the YMH theory,
\begin{eqnarray}
\theta_{\mu\nu} &=& F_{\mu}^{a \beta} F^a_{\nu \beta} - \frac{1}{4}g_{\mu\nu}F^a_{\alpha\beta}F^{a\alpha\beta} + D_\mu\Phi^a D_\nu\Phi^a\nonumber\\ 
&-& \frac{1}{2}g_{\mu\nu}(D_\alpha \Phi^a D^\alpha \Phi^a + \frac{1}{4}\lambda(\Phi^a \Phi^a - \xi^2)^2),
\label{eq.30}
\end{eqnarray}

\noindent and some calculations as shown in Ref. \cite{kn:16} - \cite{kn:18}, the total angular momentum in unit of $4\pi\xi$ is found to be
\begin{eqnarray}
J_z = \frac{1}{2\xi}\lim_{r\rightarrow \infty}r^2\partial_r\tau(r),
\label{eq.31}
\end{eqnarray}
if we assume that the time component of gauge field, $A^a_0$, is parallel to the Higgs field, $\Phi^a$, at spatial infinity.
Hence $J_z = \frac{1}{2}Q$ and the new dyon solutions possess kinetic energy of rotation. 



In the electrically charged BPS limit when the Higgs potential vanishes, the energy which is a minimum is given by \cite{kn:15}
\begin{equation}
E_{min} = \frac{4\pi\xi}{g}\sqrt{M_H^2 + Q^2},
\label{eq.32}
\end{equation}
where $M_H$ is the ``topological magnetic charge" and $Q$ as given by Eq. (\ref{eq.26}) is the total electric charge of the system when the vacuum expectation value of the Higgs field, $\xi$, is non zero.
Obviously the new dyon solution is a non BPS solution even in the limit of vanishing $\lambda$, hence its energy must be greater than that given by Eq. (\ref{eq.32}). Its dimensionless value is given by
\begin{eqnarray}
E=\frac{g}{4\xi}\int{\left\{B^a_iB^a_i + E^a_iE^a_i + D_i\Phi^aD_i\Phi^a + D_0\Phi^aD_0\Phi^a + \frac{\lambda}{2}(\Phi^a\Phi^a-\xi^2)^2\right\}d^3x}.
\label{eq.33}
\end{eqnarray}
Since this dyon solution possesses an integrable singular energy density, we define the weighted energy density to be 
\begin{eqnarray}
{\cal E}=\mbox{dimensionless energy density} \times 2\pi r^2 \sin\theta
\label{eq.34}
\end{eqnarray}
which can be plotted as shown in Figure \ref{fig.3} (b).


\section{The Dyon Solution}
\label{sect:4}

\subsection{The Numerical Construction}
\label{sect:4.1}

The numerical one and a half dyon solution was solved by using the ansatz (\ref{eq.11}) which reduced the equations of motion into eight coupled nonlinear second order partial differential equations. 
This new dyon solution is constructed by modifying the exact one and a half monopole solution of Ref. \cite{kn:17} to include the electric charge parameter, $0\leq\eta\leq 1$,
\begin{eqnarray}
\psi_1&=& \frac{3}{2}, ~~~P_1 = \sin\theta + \frac{1}{2}\sin \frac{1}{2}\theta (1+\cos\theta), \nonumber\\
R_1&=&0, ~~~P_2=\cos\theta - \frac{1}{2}\cos \frac{1}{2}\theta(1+\cos\theta), \nonumber\\
\tau_1&=& \eta\xi \cos \frac{1}{2}\theta,~~~\tau_2=\eta\xi \sin \frac{1}{2}\theta, \nonumber\\
\Phi_1&=& \xi \cos \frac{1}{2}\theta,~~~\Phi_2=\xi \sin \frac{1}{2}\theta.
\label{eq.35}
\end{eqnarray}
and using it as asymptotic solution at large distances ($r\rightarrow\infty$). In this asymptotic region, the time component of the gauge field and the Higgs field are assumed to be parallel in the isospin space, that is $\Phi_1 \propto \tau_1$ and $\Phi_2 \propto \tau_2$. However this is not necessarily true at finite $r$.

Near the origin, $r=0$, we have the common trivial vacuum solution. The asymptotic solution and boundary conditions at small distances that will give rise to finite energy solution are
\begin{eqnarray}
\psi_1=P_1=R_1=P_2=0,~~~\Phi_1=\xi_0\cos\theta, ~~~\Phi_2=-\xi_0\sin\theta,
\label{eq.36}\\
\sin\theta\tau_1(0,\theta)+\cos\theta\tau_2(0,\theta)=0,\nonumber\\
\sin\theta\Phi_1(0,\theta)+\cos\theta\Phi_2(0,\theta)=0,\nonumber\\
\partial_r(\cos\theta\tau_1(r,\theta)-\sin\theta\tau_2(r,\theta))|_{r=0}=0,\nonumber\\
\partial_r(\cos\theta\Phi_1(r,\theta)-\sin\theta\Phi_2(r,\theta))|_{r=0}=0.
\label{eq.37}
\end{eqnarray}

The boundary conditions imposed along the positive $z$-axis for the profile functions (\ref{eq.11}) of the dyon solution are
\begin{eqnarray}
\partial_\theta \Phi_1(r,\theta)|_{\theta=0} = 0, ~~\Phi_2(r,0)=0,~~ \partial_\theta \tau_1(r,\theta)|_{\theta=0} = 0, ~~\tau_2(r,0)=0,\nonumber\\
\partial_\theta \psi_1(r,\theta)|_{\theta=0} = 0, ~~R_1(r,0)=0, ~~P_1(r,0)=0, ~~\partial_\theta P_2(r,\theta)|_{\theta=0}=0,
\label{eq.38}
\end{eqnarray}
and along the negative $z$-axis, the boundary conditions imposed are
\begin{eqnarray}
\Phi_1(r,\pi)=0,~~ \partial_\theta \Phi_2(r,\theta)|_{\theta=\pi} = 0, ~~\tau_1(r,\pi)=0,~~ \partial_\theta \tau_2(r,\theta)|_{\theta=\pi} = 0,\nonumber\\
\partial_\theta \psi_1(r,\theta)|_{\theta=\pi} = 0, ~~R_1(r,\pi)=0, ~~P_1(r,\pi)=0, ~~\partial_\theta P_2(r,\theta)|_{\theta=\pi}=0.
\label{eq.39}
\end{eqnarray}

In our work we set the expectation value, $\xi=1$, and the gauge coupling constant, $g=1$. 
The new dyon solution was obtained numerically by connecting the exact asymptotic solution (\ref{eq.35}) at large distances to the trivial vacuum solution (\ref{eq.36}) at small distances and subjected to the boundary conditions (\ref{eq.37}) - (\ref{eq.39}) together with the gauge fixing condition \cite{kn:5}
\begin{equation}
r\partial_rR_1-\partial_\theta \psi_1=0.
\label{eq.40}
\end{equation}
The Maple and MATLAB software \cite{kn:17} are used for this calculation. Using the finite difference approximation, the eight reduced second order partial differential equations of motion are transformed into a system of nonlinear equations. The system of nonlinear equations are then discretized on a non-equidistant grid of size $90\times80$ covering the integration regions $0\leq \bar{x} \leq 1$ and $0\leq \theta \leq \pi$, where $\bar{x}=\frac{r}{r+1}$ is the finite interval compactified coordinate. The first and second order partial derivatives with respect to the $r$ are then replaced accordingly by ~$\partial_r \rightarrow (1-\bar{x})^2 \partial_{\bar{x}}$~ and ~$\frac{\partial^2}{\partial r^2} \rightarrow (1-\bar{x})^4\frac{\partial^2}{\partial \bar{x}^2} - 2(1-\bar{x})^3\frac{\partial}{\partial \bar{x}}$~. Maple was used to find the Jacobian sparsity pattern for the system of nonlinear equations, after which this information was provided to MATLAB to run the numerical computation. With suitable initial conditions, the system of nonlinear equations can be solved numerically using the trust-region-reflective algorithm.

The second order equations of motion Eq. (\ref{eq.3}) are solved when the $\phi$-winding number $n=1$, the electric charge parameter, $0\leq\eta<1$, and with Higgs potential when the Higgs self-coupling constant $0\leq\lambda\leq 12$.
The errors in this numerical computation come from the finite difference approximation of the functions which is of the order of $10^{-4}$ and also from the linearization of the nonlinear equations for MATLAB to solve numerically which is of the order of $10^{-6}$. Hence the overall error estimate is $10^{-4}$.

\subsection{The Numerical Results}
\label{sect:4.2}

\begin{table}[tbh]
\begin{center}

\begin{tabular}{|c|c|c|c|c|c|c|c|c|c|c|c|}

\hline 
  \multicolumn{12}{|c|} {  $\lambda=10.0$ \cellcolor{yellow} }  \\ 
\hline

$\eta$ & 0 & 0.05 & 0.1 & 0.15 & 0.2 & 0.40 & 0.80 & 0.90 & 0.94 & 0.99 & 1.00 \\ \hline

$Q$ & 0 & 0.04 & 0.08 & 0.12 & 0.16 & 0.33 & 0.72 & 0.84 & 0.90 & 0.97 & 0.99 \\ \hline

$\mu_m$ & 3.41 & 3.41 & 3.41 & 3.40 & 3.39 & 3.34 & 3.06 & 2.94 & 2.89 & 2.82 & 2.80 \\ \hline

$E$ & 1.87 & 1.87 & 1.87 & 1.88 & 1.88 & 1.94 & 2.19 & 2.31 & 2.36 & 2.44 & 2.45 \\ \hline

$d_z$ & 1.54 & 1.54 & 1.54 & 1.54 & 1.54 & 1.54 & 1.51 & 1.49 & 1.48 & 1.46 & 1.46 \\ \hline

\hline 
  \multicolumn{12}{|c|} {  $\lambda=1.0$ \cellcolor{yellow} }  \\ 
\hline

$\eta$ & 0 & 0.05 & 0.1 & 0.15 & 0.2 & 0.40 & 0.80 & 0.90 & 0.94 & 0.99 & 1.00 \\ \hline

$Q$ & 0 & 0.04 & 0.09 & 0.13 & 0.18 & 0.37 & 0.83 & 0.99 & 1.06 & 1.15 & 1.17 \\ \hline

$\mu_m$ & 4.01 & 4.01 & 4.00 & 4.00 & 3.99 & 3.93 & 3.64 & 3.51 & 3.44 & 3.35 & 3.33 \\ \hline

$E$ & 1.63 & 1.63 & 1.63 & 1.64 & 1.64 & 1.71 & 2.01 & 2.15 & 2.21 & 2.31 & 2.33 \\ \hline

$d_z$ & 2.08 & 2.08 & 2.09 & 2.09 & 2.09 & 2.11 & 2.16 & 2.15 & 2.14 & 2.12 & 2.12 \\ \hline

\hline 
  \multicolumn{12}{|c|} {  $\lambda=0$ \cellcolor{yellow} }  \\ 
\hline

$\eta$ & 0 & 0.05 & 0.1 & 0.15 & 0.2 & 0.40 & 0.80 & 0.90 & 0.94 & 0.99 & 1.00 \\ \hline

$Q$ & 0 & 0.06 & 0.12 & 0.19 & 0.25 & 0.54 & 1.65 & 2.54 & 3.38 & 8.20 & - \\ \hline

$\mu_m$ & 5.93 & 5.93 & 5.94 & 5.95 & 5.97 & 6.12 & 7.33 & 8.69 & 10.12 & 19.42 & - \\ \hline

$E$ & 1.23 & 1.23 & 1.24 & 1.24 & 1.26 & 1.35 & 2.08 & 2.87 & 3.68 & 8.30 & - \\ \hline

$d_z$ & 3.57 & 3.57 & 3.58 & 3.61 & 3.64 & 3.89 & 5.96 & 8.23 & 10.56 & 25.60 & - \\ \hline

\end{tabular}
\end{center}
\caption{Table of the electric charge $Q$, dimensionless magnetic dipole moment $\mu_m$, dimensionless total energy $E$ and monopoles' separation $d_z$ of the one and a half dyon for different values of $\eta$ at $\lambda=0$, $\lambda=1.0$ and $\lambda=10.0$.}
\label{table.1}
\end{table}

\begin{table}[tbh]
\begin{center}

\begin{tabular}{|c|c|c|c|c|c|c|c|c|c|c|c|}

\hline 
  \multicolumn{12}{|c|} {  $\eta=1.0$ \cellcolor{yellow} }  \\ 
\hline

$\lambda$ & 0 & 0.04 & 0.09 & 0.20 & 0.40  & 0.80 & 1.00 & 2.00 & 4.00 & 8.00 & 12.00 \\ \hline

$Q$ & - & 1.68 & 1.52 & 1.38 & 1.28 & 1.19 & 1.17 & 1.10 & 1.04 & 1.00 & 0.98 \\ \hline

$\mu_m$ & - & 5.05 & 4.52 & 4.05 & 3.70 & 3.41 & 3.33 & 3.12 & 2.96 & 2.84 & 2.78 \\ \hline

$E$ & - & 2.44 & 2.37 & 2.33 & 2.32 & 2.32 & 2.33 & 2.36 & 2.39 & 2.44 & 2.46 \\ \hline

$d_z$ & - & 3.94 & 3.36 & 2.87 & 2.51 & 2.21 & 2.12 & 1.88 & 1.68 & 1.51 & 1.42 \\ \hline

\hline 
  \multicolumn{12}{|c|} {  $\eta=0.9$ \cellcolor{yellow} }  \\ 
\hline

$\lambda$ & 0 & 0.04 & 0.09 & 0.20 & 0.40  & 0.80 & 1.00 & 2.00 & 4.00 & 8.00 & 12.00 \\ \hline

$Q$ & 2.54 & 1.35 & 1.24 & 1.15 & 1.07 & 1.01 & 0.99 & 0.93 & 0.89 & 0.85 & 0.83  \\ \hline

$\mu_m$ & 8.69 & 5.21 & 4.71 & 4.25 & 3.89 & 3.59 & 3.51 & 3.28 & 3.11 & 2.98 & 2.92 \\ \hline

$E$ & 2.87 & 2.11 & 2.09 & 2.09 & 2.11 & 2.14 & 2.15 & 2.19 & 2.24 & 2.29 & 2.32 \\ \hline

$d_z$ & 8.23 & 3.75 & 3.28 & 2.85 & 2.52 & 2.23 & 2.15 & 1.91 & 1.71 & 1.54 & 1.45 \\ \hline

\hline 
  \multicolumn{12}{|c|} {  $\eta=0.75$ \cellcolor{yellow} }  \\ 
\hline

$\lambda$ & 0 & 0.04 & 0.09 & 0.20 & 0.40  & 0.80 & 1.00 & 2.00 & 4.00 & 8.00 & 12.00 \\ \hline

$Q$ & 1.40 & 0.99 & 0.93 & 0.87 & 0.82 & 0.78 & 0.77 & 0.73 & 0.70 & 0.67 & 0.66 \\ \hline

$\mu_m$ & 7.00 & 5.29 & 4.86 & 4.44 & 4.09 & 3.78 & 3.70 & 3.47 & 3.29 & 3.15 & 3.08 \\ \hline

$E$ & 1.88 & 1.80 & 1.82 & 1.85 & 1.89 & 1.94 & 1.95 & 2.01 & 2.07 & 2.13 & 2.16 \\ \hline

$d_z$ & 5.40 & 3.47 & 3.12 & 2.77 & 2.49 & 2.23 & 2.15 & 1.93 & 1.74 & 1.57 & 1.48 \\ \hline

  \multicolumn{12}{|c|} {  $\eta=0.5$ \cellcolor{yellow} }  \\ 
\hline

$\lambda$ & 0 & 0.04 & 0.09 & 0.20 & 0.40  & 0.80 & 1.00 & 2.00 & 4.00 & 8.00 & 12.00 \\ \hline

$Q$ & 0.71 & 0.58 & 0.55 & 0.52 & 0.50 & 0.48 & 0.47 & 0.45 & 0.43 & 0.42 & 0.41  \\ \hline

$\mu_m$ & 6.25 & 5.30 & 4.96 & 4.59 & 4.27 & 3.97 & 3.89 & 3.66 & 3.47 & 3.33 & 3.26 \\ \hline

$E$ & 1.43 & 1.53 & 1.57 & 1.62 & 1.68 & 1.74 & 1.76 & 1.82 & 1.89 & 1.96 & 2.00 \\ \hline

$d_z$ & 4.12 & 3.14 & 2.90 & 2.64 & 2.41 & 2.19 & 2.12 & 1.92 & 1.75 & 1.59 & 1.50 \\ \hline

\end{tabular}
\end{center}
\caption{Table of the electric charge $Q$, dimensionless magnetic dipole moment $\mu_m$, dimensionless total energy $E$ and the monopoles' separation $d_z$ of the one and a half dyon for different values of $\lambda$ at $\eta$ = 0.5, 0.9, 0.75 and 1.0.}
\label{table.2}
\end{table}

\begin{figure}[tbh]
	\centering
	\hskip-0.1in
	 \includegraphics[width=6.0in,height=6.5in]{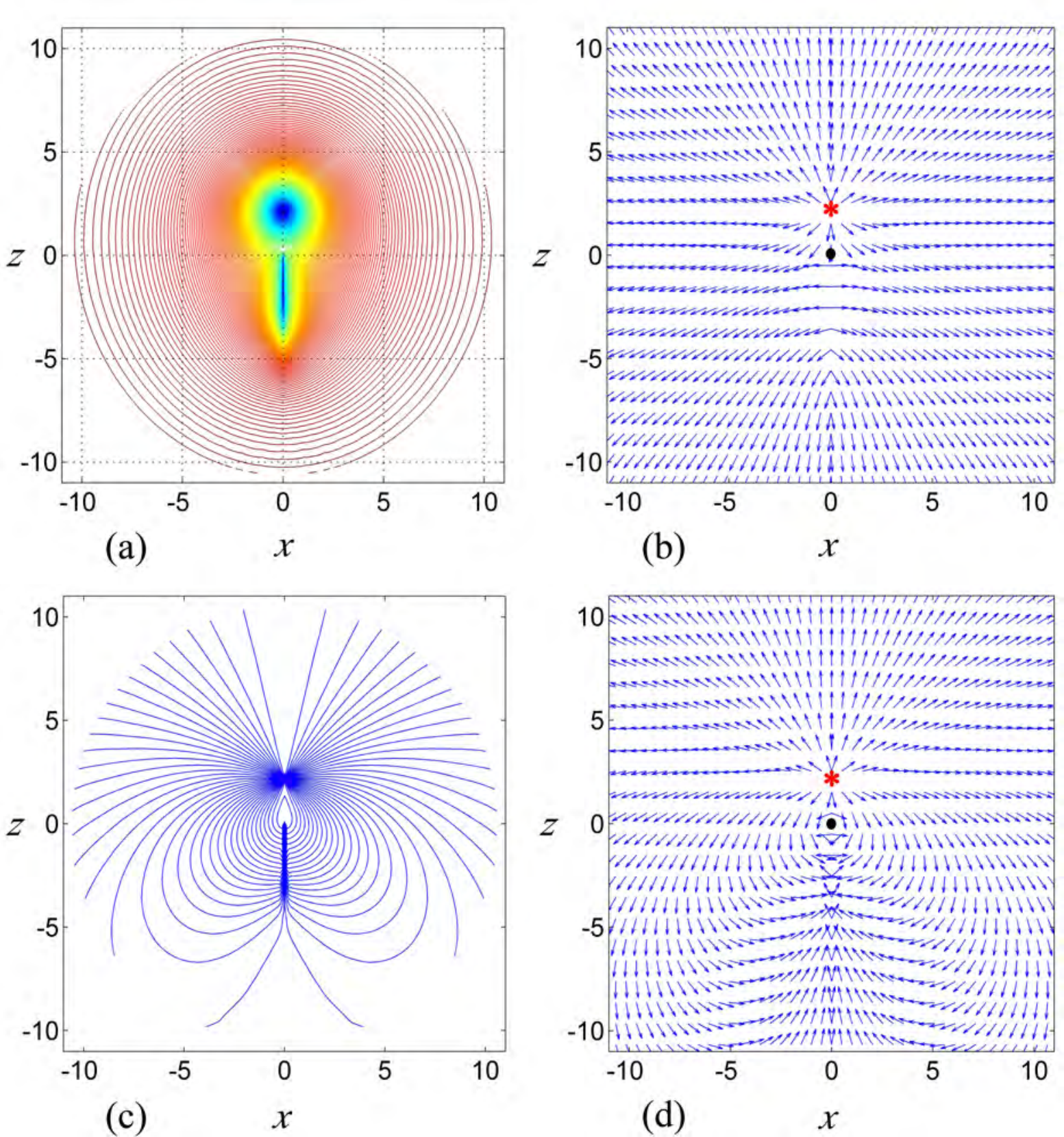} 
	\caption{(a) Contour plot of the electric field equipotential lines. (b) Vector field plot of the electric field unit vector, $\hat{E}_i$. (c) Contour plot of the magnetic field lines. (d) Vector field plot of the magnetic field unit vector, $\hat{B}_i$. Here  $\lambda=1$ and $\eta=0.9$.}
	\label{fig.1}
\end{figure}

\begin{figure}[tbh]
	\centering
	\hskip0in
	 \includegraphics[width=6.0in,height=7.0in]{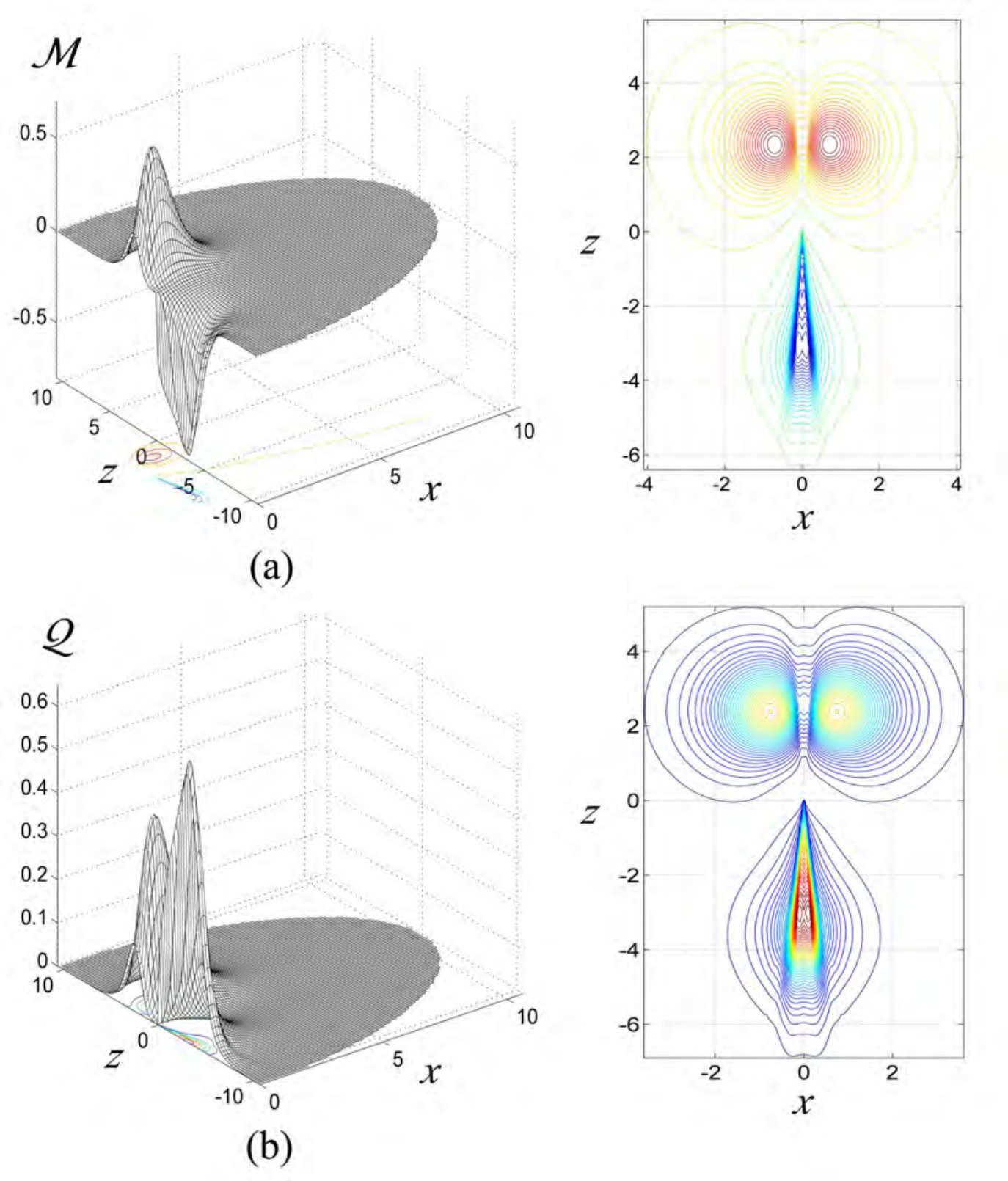} 
	\caption{The 3D surface and contour plots of (a) the weighted magnetic charge density, ${\cal M}$, and (b) the weighted electric charge density, ${\cal Q}$ of the one and a half dyons solution along the $x$-$z$ plane at $y=0$ when $\lambda=1$ and $\eta=0.9$.}
	\label{fig.2}
\end{figure}

\begin{figure}[tbh]
	\centering
	\hskip-0.1in
	 \includegraphics[width=5.5in,height=7.5in]{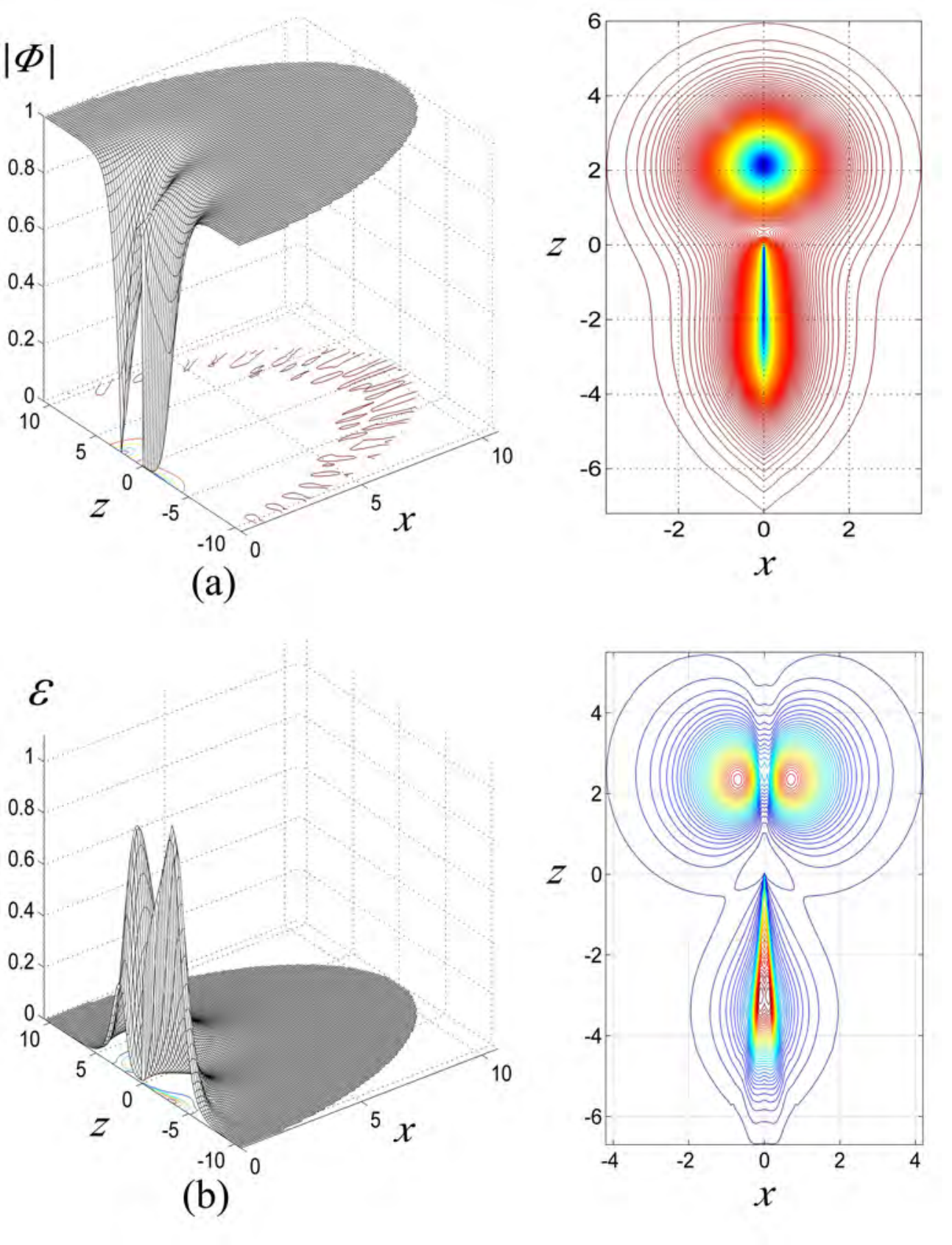} 
	\caption{The 3D surface and contour plots of (a) the Higgs field modulus, $|\Phi|$, and (b) the weighted energy density, ${\cal E}$, of the one and a half dyons solution along the $x$-$z$ plane at $y=0$ when $\lambda=1$ and $\eta=0.9$.}
	\label{fig.3}
\end{figure}

The numerical solutions obtained for the new dyon solution are all regular functions of $r$ and $\theta$ except for the profile function $R_2$ which is singular along the $z$-axis. 
We calculate and draw the 3D surface graphs together with their respective contour plots for the Higgs field modulus $|\Phi|$ (Figure \ref{fig.3} (a)), the weighted energy density ${\cal E}$ (Figure \ref{fig.3} (b)), the weighted magnetic charge density ${\cal M}$ (Figure \ref{fig.2} (a)), and the weighted electric charge density ${\cal Q}$ (Figure \ref{fig.2} (b)), versus the $x$-$z$ plane at $y=0$ numerically. The 3D surface plot of these physical quantities and their respective contour plot are drawn for the case when $\lambda=1$ and $\eta=0.9$. The magnetic charge density is positve above the $x$-$y$ plane and negative below it. Hence the 't Hooft-Polyakov monopole located along the positive $z$-axis possesses magnetic charge $+1$ and the one-half monopole located at $r=0$ possesses magnetic charge $-\frac{1}{2}$. The electric charge density is however positive throughout space and this is so when the electric charge parameter $\eta$ is positive. 
Hence from Figure \ref{fig.2} we can conclude that the one-half dyon carries negative magnetic and positive electric charge densities that are concentrated along a finite stretch of the negative $z$-axis near the origin while the 't Hooft-Polyakov monopole at $z=2.15$ carries both positive magnetic and electric charge densities that are spherically concentrated around the monopole. The sign of the electric and magnetic charges are once again confirm by the contour plots of the electric field equipotential lines and the magnetic field lines as shown in Figure \ref{fig.1} (a) and (c) respectively and the vector field plots of the electric field unit vector, $\hat{E}_i$, and the magnetic field unit vector, $\hat{B}_i$ as shown in Figure \ref{fig.1} (b) and (d) respectively. 

\begin{figure}[tbh]
	\centering
	\hskip-0.1in
	 \includegraphics[width=5.5in,height=6in]{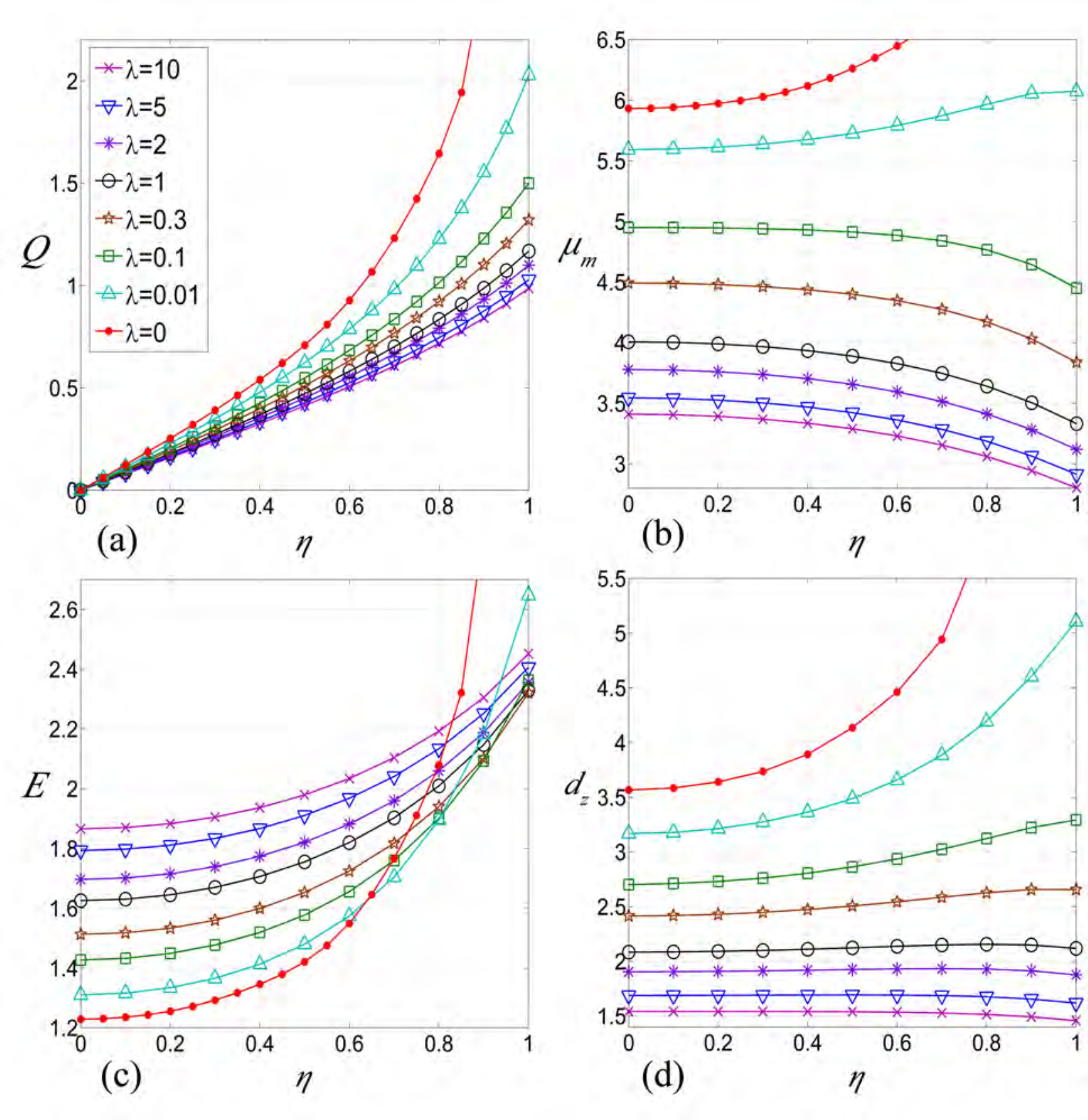} 
	\caption{(a) Plots of the total electric charge, $Q$, (b) the magnetic dipole moment, $\mu_m$, (c) the total energy, $E$, and (d) the poles' separation, $d_z$, versus $\eta$. Here, $0\leq\lambda\leq 10$, and, $0\leq \eta<1$.}
	\label{fig.4}
\end{figure}

In general, the shape of the dyon remains the same as that of the one and a half monopole solution of Ref. \cite{kn:10} while its size increases as $\eta$ increases from zero to one. This is because as the electric charge parameter $\eta\rightarrow 1$, the zeros of the Higgs modulus along the negative $z$-axis increase and the inverted cone becomes more stretched along the negative $z$-axis. However at $\lambda=1$, the poles' separation, $d_z$, does not varied much with $\eta$ as shown in the graphs of Figure \ref{fig.4} (d).

The quantities that vary with the electric charge parameter, $\eta$, are the total charge $Q$, the magnetic dipole moment, $\mu_m$, the total energy, $E$, and the dipole separation, $d_z$. The graphs for eight different values of $\lambda=0, 0.01, 0.1, 0.3, 1, 2, 5$, and 10 for each of the four quantities versus $\eta$ are shown in Figure \ref{fig.4}. The numerical values are given in Table \ref{table.1} for $\lambda= 0, 1, 10$. From Figure \ref{fig.4} (a), the graphs of $Q$ versus $\eta$ are all nondecreasing graphs with the critical value of $Q_{critical}$ increasing with decreasing $\lambda$. When $\lambda\rightarrow0$, $Q_{critical}\rightarrow\infty$ and this behaviour is similar to the other dyons solutions of Ref. \cite{kn:15} - \cite{kn:18}.

The graphs of total energy, $E$, versus $\eta$, Figure \ref{fig.4} (c), are similar to the graphs of Figure \ref{fig.4} (a), that is $Q$ versus $\eta$, as they are all nondecreasing graphs. However the critical value of $E$ for large values of $\lambda$ decreases with decreasing $\lambda$ from $\lambda=12$ ($E_{critical}=2.4646$) to $\lambda=0.4$ ($E_{critical}=2.3183$) after which it increases to infinity as $\lambda$ approaches zero. Similar to the dyon solutions of Ref. \cite{kn:15} - \cite{kn:18} for small $\lambda<0.4$, $E_{critical}\rightarrow\infty$ as $\lambda\rightarrow0$.

The graphs of the magnetic dipole moment, $\mu_m$, versus $\eta$, Figure \ref{fig.4} (b), and the dipole separation, $d_z$, versus $\eta$, Figure \ref{fig.4} (d), possess critical values that increase with decreasing $\lambda$ such that as $\lambda\rightarrow0$, ~~$\{\mu_m^{critical}, ~d_z^{critical}\}\rightarrow\infty$. This behaviour is common to the other dyon solutions. However the graphs of Figure \ref{fig.4} (b) and (d) are nondecreasing graphs only for small $\lambda\leq 0.01$. For larger values of $\lambda$, the graphs become nonincreasing and $\mu_m$ and $d_z$ decrease with $\eta$ which differ from the norm.

\begin{figure}[tbh]
	\centering
	\hskip0in
	 \includegraphics[width=5.5in,height=8.0in]{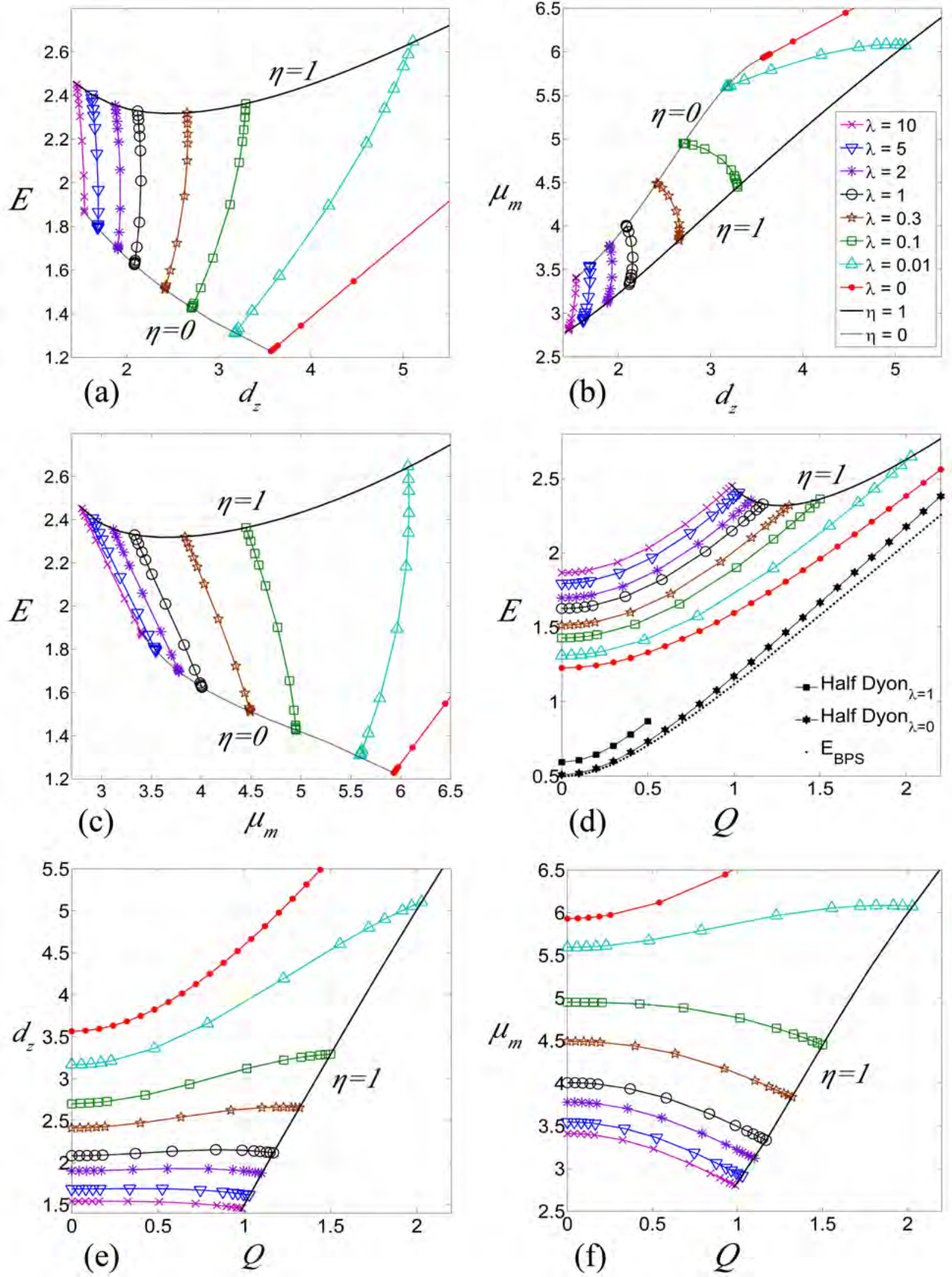} 
	\caption{Plots of (a) the total energy, $E$, and (b) magnetic dipole moment, $\mu_m$, versus the poles' separation, $d_z$. (c) Plots of $E$ versus $\mu_m$.  Plots of (d) $E$, (e) $d_z$, and (f) $\mu_m$, versus the electric charge, $Q$. Here, $0\leq\lambda\leq 10$, and, graphs of constant $\eta=0$ and 1 are shown.}
	\label{fig.5}
\end{figure}

In Figure \ref{fig.5}, the graphs of the four quantities, the total charge $Q$, the magnetic dipole moment, $\mu_m$, the total energy, $E$, and the dipole separation, $d_z$, are plotted versus each other for constant values of $\lambda= 0, 0.01, 0.1, 0.3, 1, 2, 5, 10$ and for $\eta=0, 1$. The graphs of magnetic dipole moment $\mu_m$ versus $d_z$, Figure \ref{fig.5} (b); dipole separation $d_z$ versus $Q$, Figure \ref{fig.5} (e); and magnetic dipole moment $\mu_m$, versus $Q$, Figure \ref{fig.5} (f); at $\eta=1$ are almost linear. The strange behaviour of $\mu_m$ and $d_z$ decreasing with increasing $\eta$ at large values of $\lambda$ are once again reflected in the graphs of Figure \ref{fig.5} (a), (b), (c), (e) and (f). The graphs of total energy $E$ versus $Q$, Figure \ref{fig.5} (d), for constant values of $\lambda$ however behave normally as they are all nondecreasing graphs. 

\begin{figure}[tbh]
	\centering
	\hskip0in
	 \includegraphics[width=5.5in,height=8.0in]{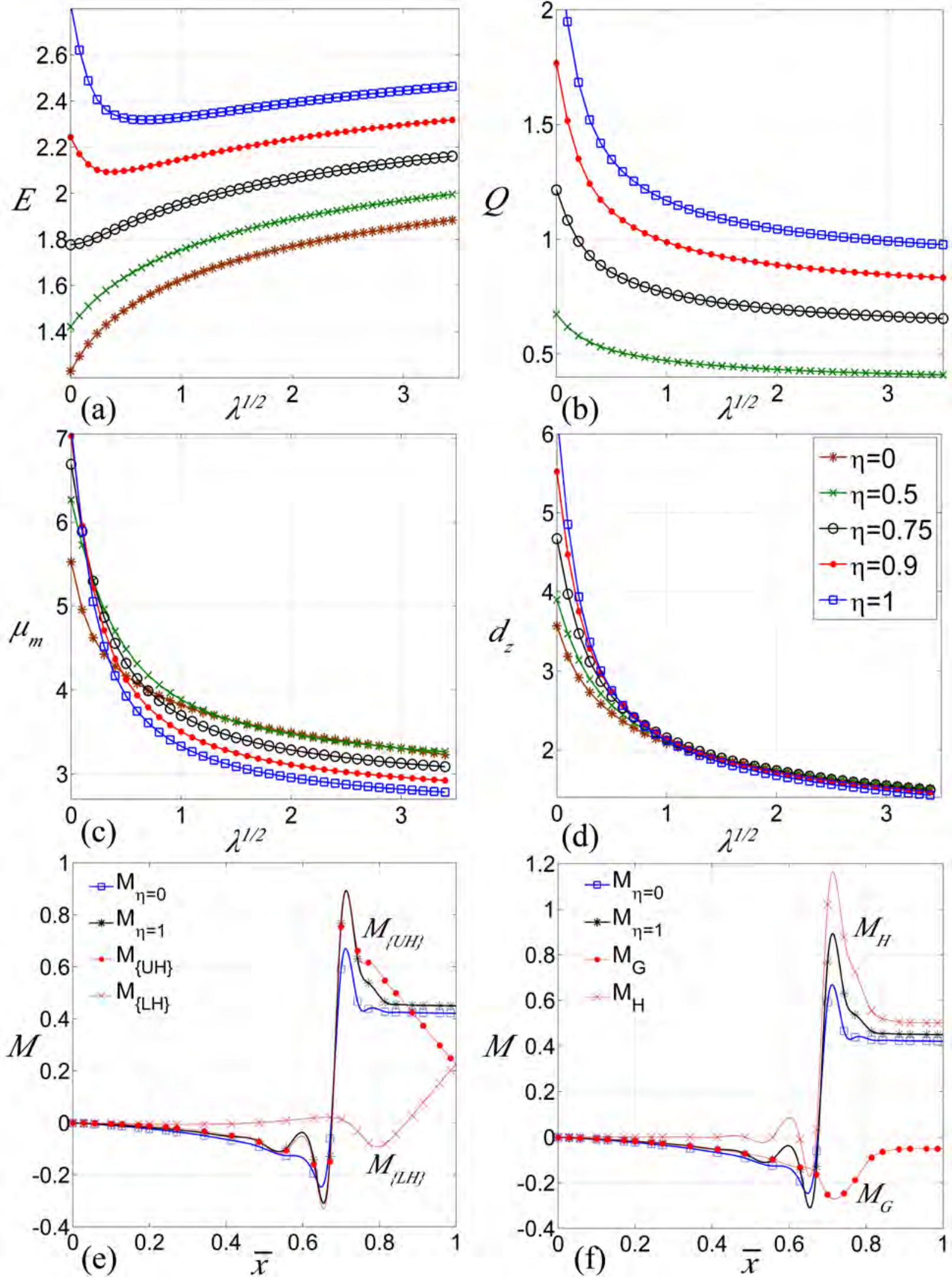} 
	\caption{(a) Plot of the total energy, $E$, (b) the total electric charge, $Q$, (c) the magnetic dipole moment, $\mu_m$, and (d) the poles' separation, $d_z$, versus $\sqrt{\lambda}$ for $\eta=0, 0.50, 0.75, 0.90$ and 1. Plots of (e) magnetic charges, $M$, $M_{UH}$, and $M_{LH}$ and (f) $M$, $M_H$, and $M_G$, versus the compactified coordinate, $\bar{x}$ when $\lambda=1$ and $\eta= 1$.}
	\label{fig.6}
\end{figure}

The total energy, $E$, the total charge $Q$, the magnetic dipole moment, $\mu_m$, and the dipole separation, $d_z$ are plotted versus $\lambda^{1/2}$ in Figure \ref{fig.6} (a) to (d) respectively. The behaviour of the graphs in Figure \ref{fig.6} (a) to (c) are similar to that of the one-half dyon solution \cite{kn:17} as $E$ increases with $\lambda^{1/2}$ for values of $\lambda\geq 1$ and the graphs of $Q$ and $\mu_m$ versus $\lambda^{1/2}$ are all nonincreasing graphs. The graphs of $d_z$ versus $\lambda^{1/2}$ in Figure \ref{fig.6} (d) are also nonincreasing graphs. At small values of, $0<\lambda^{1/2}<0.548$, the dipole separation, $d_z$, increases with increasing $\eta$, which is an expected behaviour due to increase electrical repulsion. However when $\lambda^{1/2}> 2.746$, the dipole separation, $d_z$ decreases with increasing $\eta$.

The graphs of the magnetic charges, $M_{\{UH\}}$, $M_{\{LH\}}$, $M_H$, $M_G$, and $M$ when $\lambda=\eta=1$ are shown in Figure \ref{fig.6} (e) and (f). The fact that $M_{\{UH\}}=M_{\{LH\}}$ at $\bar{x}=1$ or $r=\infty$ shows that the magnetic field is radially symmetrical at large distances. The value of $M_H$ is zero at $r=0$ and $M_H=0.50$ at $r=\infty$. This indicates that the net topological charge of the system is zero. The topological charge of $-0.5$ carried by the Dirac string at $r$ infinity cannot be captured by the numerical calculation. The region between the over-shoot and under-shoot of the graph for $M_H$ is the region where the 't Hooft-Polyakov monopole is located. The graphs also show that the 't Hooft-Polyakov monopole is a real particle of magnetic charge $M=1$ whereas the one-half monopole of magnetic charge $M=-\frac{1}{2}$ is a virtual particle.


\subsection{The Cho Decomposition of the Dyon Solution}
\label{sect:4.3}

In the Cho decomposition of the SU(2) YMH theory, the ansatz that follows is gauge independent once the Higgs field direction, $\hat{\Phi}^a$, is chosen \cite{kn:19}. The magnetic ansatz (\ref{eq.11}) used to construct the solution is not gauge independent and the gauge independent ansatz can be given by the Cho decomposition. The Higgs field direction can be defined by writing the axially symmetric Higgs unit vector, $\hat{\Phi}^a=h_1\hat{n}^a_r+h_2\hat{n}^a_\theta$, in the rectangular coordinate system as in Eq. (\ref{eq.16}) and (\ref{eq.17}).
Hence the first and second perpendicular of the Higgs field unit vector in isospin space are,

\begin{eqnarray}
&&\hat{\Phi}_1^a = -h_2\hat{n}^a_r+h_1\hat{n}^a_\theta=\cos\alpha \cos n\phi ~\delta_{1}^a + \cos\alpha \sin n\phi ~\delta_{2}^a - \sin\alpha \delta_{3}^a,\nonumber\\
&&\hat{\Phi}_2^a = \hat{n}^a_\phi=-\sin n\phi ~\delta_{1}^a + \cos n\phi ~\delta_{2}^a.
\label{eq.41}
\end{eqnarray}

The Cho decomposition of the YM gauge potential is given by Ref. \cite{kn:19} as,
\begin{eqnarray}
A_\mu^a &=& \hat{A}^a_\mu + X^a_\mu, ~~X^a_\mu\hat{\Phi}^a=0,
\label{eq.42}\\
\mbox{where} ~~\hat{A}^a_\mu &=& A_\mu\hat{\Phi}^a - \frac{1}{g}\epsilon^{abc}\hat{\Phi}^b\partial_\mu\hat{\Phi}^c,
\label{eq.43}
\end{eqnarray}
is the restricted gauge potential and $X^a_\mu$ is the valence gauge potential. Both the Higgs part of the gauge potential, ~$- \frac{1}{g}\epsilon^{abc}\hat{\Phi}^b\partial_\mu\hat{\Phi}^c$~ and the non-Abelian gauge potential, $X^a_\mu$ are perpendicular to the Higgs field direction in isospin space, $\hat{\Phi}^a$. 
The gauge independent Cho decomposition axially symmetric ansatz then takes the form \cite{kn:11},
\begin{eqnarray}
gA_i^a &=& A ~\hat{\phi}_i\hat{\Phi}^a + (X_1+Y_1)\hat{\phi}_i\hat{\Phi}_1^a + (X_3+Y_3)\hat{r}_i\hat{\Phi}_2^a+(X_4+Y_4)\hat{\theta}_i\hat{\Phi}_2^a, \nonumber\\
gA^a_0 &=& A_0\hat{\Phi}^a + X_0\hat{\Phi}_1^a, ~~~g\Phi^a = |\Phi|~\hat{\Phi}^a,
\label{eq.44}
\end{eqnarray}
where $X_0$, $X_1$, $X_3$, $X_4$ are the component of the gauge covariant valence potential $X^a_\mu$ and $Y_1$, $Y_3$, $Y_4$, $A_0$, $A$ are the component of the restricted gauge potential $\hat{A}^a_\mu$. Comparing the ansatz (\ref{eq.44}) with the magnetic ansatz (\ref{eq.11}), we note that,
\begin{eqnarray}
&&X_0 = \tau_2 h_1-\tau_1 h_2, ~~X_1+Y_1 = \frac{1}{r}(h_1\psi_2+h_2 R_2),\nonumber\\
&&X_3+Y_3 = \frac{1}{r}R_1, ~~X_4+Y_4 = - \frac{1}{r}\psi_1 \nonumber\\
&&A_0 = \tau_1 h_1+\tau_2 h_2, ~~A = \frac{1}{r}(h_2\psi_2-h_1 R_2), ~~|\Phi| = \sqrt{\Phi_1^2+\Phi_2^2},
\label{eq.45}
\end{eqnarray}
where
\begin{eqnarray}
Y_1 &=& \frac{n}{r\sin\theta}\left(h_1\sin\theta + h_2\cos\theta\right) = \frac{n\sin\alpha}{r\sin\theta},  \nonumber\\
Y_3 &=& \frac{\partial_r h_1}{h_2} = - \partial_r \alpha, ~~Y_4 = -\frac{1}{r}\left(1-\frac{\partial_\theta h_1}{h_2}\right) = - \frac{1}{r}\partial_\theta \alpha.
\label{eq.46}
\end{eqnarray}

The electromagnetic field strength tensor of the Cho decomposed gauge potential (\ref{eq.42}) is given by
\begin{eqnarray}
F^a_{\mu\nu} = \hat{F}_{\mu\nu}^a + \hat{D}_\mu X^a_\nu - \hat{D}_\nu X^a_\mu + g\epsilon^{abc}X_\mu^b X_\nu^c,
\label{eq.47}
\end{eqnarray}
where $\hat{F}_{\mu\nu}^a = \hat{F}_{\mu\nu}\hat{\Phi}^a$, is the field strength of the self-dual potential (\ref{eq.43}) and $\hat{F}_{\mu\nu}$ is the 't Hooft's electromagnetic field. The Higgs unit vector is invariant under the covariant derivative $\hat{D}_\mu$ of the self-dual gauge potential $\hat{A}^a_\mu$,
\begin{eqnarray}
\hat{D}_\mu\hat{\Phi}^a = \partial_\mu\hat{\Phi}^a + g\epsilon^{abc}\hat{A}^b_{\mu}\hat{\Phi}^c = 0.
\label{eq.48}
\end{eqnarray}

\begin{figure}[tbh]
	\centering
	\hskip0in
	 \includegraphics[width=6.0in,height=8.0in]{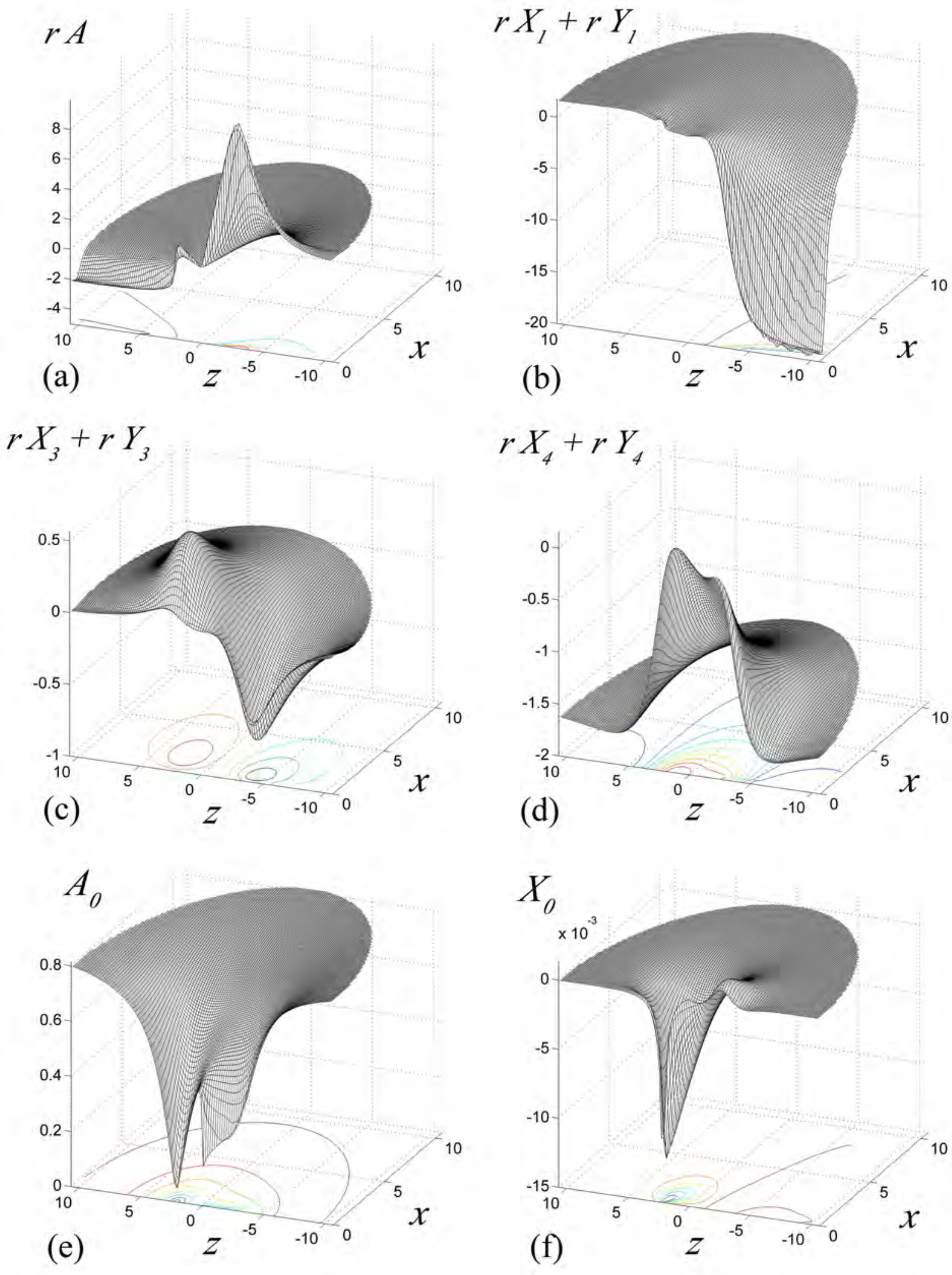} 
	\caption{3D graphs of the profile functions (a) $rA$, (b) $rX_1 + rY_1$, (c) $rX_3 + rY_3$, (d) $rX_4 + rY_4$, (e) $A_0$, and (f) $X_0$ of the one and a half dyons solution versus $x$-$z$ plane at $y=0$ for $g=\xi=\lambda=\eta=1$.}
	\label{fig.7}
\end{figure}

The numerically Cho decomposed gauge field profile functions (\ref{eq.45}) are shown in Figure \ref{fig.7} as 3D surface plots versus the $x$-$z$ plane at $y$=0. The presence of the electrically charge 't Hooft-Polyakov monopole and the one-half monopole are reflected in all the six profile functions as shown in Figure \ref{fig.7} by the curvature of the surface plots. All the profile functions are regular bounded surfaces except for $r(X_1+Y_1)$ which is singular along the negative $z$-axis, Figure \ref{fig.7} (b). The nonvanishing of the function, $X_0$, at finite values of $r$ indicates that the time component of the gauge function, $A_0^a$, is only parallel to the Higgs field in isospin space at large distances. 

\section{Comments}

Similar to the one-half dyon solution of Ref. \cite{kn:17}, the new dyon solution possesses a net magnetic charge of $M=\frac{1}{2}$ when the magnetic charge carried by the Dirac string is excluded. It also possesses nonvanishing magnetic dipole moment and angular momentum, $J_z=\frac{1}{2}Q$ when the electric charge $Q\not=0$. Hence both these dyon solutions are able to rotate in the presence of an external magnetic field. This is possible because the gauge potentials of both dyon solutions possess a semi-infinite Dirac string. 

Upon performing the Cho decomposition of the dyon solution here, the infinite string singularity along the $z$-axis can be transformed into a semi-infinite Dirac string. A more comprehensive discussion of the Cho decomposition of the electrically neutral solution is given in Ref. \cite{kn:11}.

Similar to the dyons solutions of Ref. \cite{kn:15} and \cite{kn:18}, the total electric charge, magnetic dipole moment, total energy, and dipole separation of the new dyon solution increases indefinitely as the electric charge parameter, $\eta\rightarrow 1$ when $\lambda$ vanishes. However when $\lambda$ is nonvanishing, $Q$, $\mu_m$, $E$, and $d_z$ approach critical values as given in Table {\ref{table.1} and {\ref{table.2}. The difference between this new dyon solution and the other dipole dyon solutions of Ref. \cite{kn:16} and \cite{kn:17} is that at large values of $\lambda$, the dipole separation, $d_z$, and the magnetic dipole moment, $\mu_m$, decrease instead of increase with increasing $\eta$. This behaviour is not normal because as $\eta$ increases the electric charge $Q$ also increases and this will lead to repulsion between the two poles instead of attraction. This mean that $\mu_m$ and $d_z$ should increase with increasing $\eta$. However this new dyon solution behave in the reverse way when $\lambda>1$.


\section{Acknowledgements}
The authors would like to thank Universiti Sains Malaysia for the RU research grant (account number: 1001/PFIZIK/811180) and the Ministry of Science, Technology and Innovation for the ScienceFund Grant (account number: 305/PFIZIK/613613).



\begin{thebibliography}{99}

\bibitem[1]{kn:1} G. 't Hooft, Nucl. Phy. {\bf B79},  276 (1974).

\bibitem[2]{kn:2} A.M. Polyakov, Sov. Phys. - JETP {\bf 41}, 988 (1975); A.M. Polyakov, Phys. Lett. {\bf B59},  82 (1975); A.M. Polyakov, JETP Lett. {\bf 20}, 194 (1974);  E.B. Bogomol'nyi and M.S. Marinov, Sov. J. Nucl. Phys. {\bf 23}, 355 (1976); E.B. Bogomol'nyi, Sov. J. Nucl. Phys. {\bf 24}, 449 (1976).

\bibitem[3]{kn:3} C. Rebbi and P. Rossi, Phys. Rev. {\bf D22}, 2010 (1980); R.S. Ward, Commun. Math. Phys. {\bf 79}, 317 (1981); P. Forg\'{a}cs, Z. Horv\'{a}th and L. Palla, Phys. Lett. {\bf B99}, 232 (1981); P. Forg\'{a}cs, Z. Horv\'{a}th and L. Palla, Nucl. Phys. {\bf B192}, 141 (1981); M.K. Prasad, Commun. Math. Phys. {\bf 80}, 137 (1981); M.K. Prasad and P. Rossi, Phys. Rev. {\bf D24}, 2182 (1981); Rosy Teh and K.M. Wong, J. Math. Phys. 46, 082301 (2005); Int. J. Mod. Phys. A 20, 4291 (2005).

\bibitem[4]{kn:4} P.M. Sutcliffe, Int. J. Mod. Phys. A 12 (1997) 4663; C.J. Houghton, N.S. Manton and P.M. Sutcliffe, Nucl.Phys. B 510 (1998) 507.


\bibitem[5]{kn:5} B. Kleihaus and J. Kunz, Phys. Rev. {\bf D 61}, 025003 (1999); B. Kleihaus, J. Kunz, and Y. Shnir, Phys. Lett. {\bf B570}, 237 (2003); Phys. Rev. {\bf D 68}, 101701 (2003); Phys. Rev. {\bf D 70}, 065010 (2004); J. Kunz, U. Neemann and Y. Shnir, Phys. Lett. {\bf B640}, 57 (2006).

\bibitem[6]{kn:6} Rosy Teh, K.M. Wong and K.G. Lim, Int. J. Mod. Phys. {\bf A25}, 5731 (2010); Rosy Teh, P.Y. Tan, and K.M. Wong, J. Mod. Phys {\bf A27}, 1250148, (2012).

\bibitem[7]{kn:7} E. Harikumar, I. Mitra, and H.S. Sharatchandra, Phys. Lett. B 557 (2003) 303.

\bibitem[8]{kn:8} Rosy Teh and K.M. Wong, {\it Half-Monopole and Multimonopole}, Int. J. Mod. Phys. A 20, (2005) 2195;  Rosy Teh, K.G. Lim and P.W. Koh, {\it Magnetic Half-Monopole Solutions}, FRONTIERS IN PHYSICS: 3rd International Meeting, Kuala Lumpur (Malaysia), 12-16 January 2009, edited by S.P. Chia, M.R. Muhammad, and K. Ratnavelu,  ISBN: 978-0-7354-0687-2, AIP Conference Proceedings Volume 1150, 424 (2009).

\bibitem[9]{kn:9} Rosy Teh, B.L. Ng, and K.M. Wong, {\it Finite Energy One-Half Monopole Solutions of the SU(2) Yang-Mills-Higgs Theory}, Proceedings of Science, POS (ICHEP 2012) 473; Mod. Phys. Letts. {\bf A27}, 1250233 (2012)

\bibitem[10]{kn:10} Rosy Teh, B.L. Ng, and K.M. Wong, Ann. Phys. {\bf 343C}, 1 (2014).

\bibitem[11]{kn:11} Rosy Teh, B.L. Ng, and K.M. Wong, Int. J. Mod. Phys. {\bf A28}, 1350144 (2013).

\bibitem[12]{kn:12} E. Witten, Phys. Lett. {\bf B86},  283 (1979).

\bibitem[13]{kn:13} B. Julia and A. Zee, Phys. Rev. {\bf D11}, 2227 (1975);  M.K. Prasad and C.M. Sommerfield, Phys. Rev. Lett. {\bf 35}, 760 (1975); F.A. Bais and J.R. Primack, Phys. Rev. {\bf D13}, 819 (1976). 

\bibitem[14]{kn:14} S. Coleman, S.Parke, A. Neveu, and C.M. Sommerfield, Phys. Rev. {\bf D15}, 544 (1977).

\bibitem[15]{kn:15} B. Hartmann, B. Kleihaus, and J. Kunz, Mod. Phys. Letts. {\bf A15}, 1003 (2000); Y. Brihaye, B. Kleihaus, and D.H. Tchrakian, J. Math. Phys. {\bf 40}, 1136 (1999).

\bibitem[16]{kn:16} K.G. Lim, Rosy Teh and K.M. Wong, J. Phys. G: Nucl. Part. Phys. 39 (2012) 025002.

\bibitem[17]{kn:17} Rosy Teh, B.L. Ng, and K.M. Wong, J. Phys. G: Nucl. Part. Phys. 40, 035007 (2013).

\bibitem[18]{kn:18} J.J. Van Der Bij and E. Radu, Int. J. Mod. Phys. {\bf A17}, 1477 (2002); Int. J. Mod. Phys. {\bf A18}, 2379 (2003); V. Paturyan, E. Radu and D.H. Tchrakian, Phys. Lett. {\bf B609}, 360 (2005); B. Kleihaus, J. Kunz and U. Neemann, Phys. Lett. {\bf B623}, 171 (2005).

\bibitem[19]{kn:19} Y.M. Cho, Phys. Rev. {\bf D21}, 1080 (1980); Phys. Rev. Lett. {\bf 46}, 302 (1981); Phys. Rev. {\bf D23}, 2415 (1981).

\bibitem[20]{kn:20} N.S. Manton, Nucl. Phys. (N.Y.) {\bf B126}, 525 (1977).

\bibitem[21]{kn:21} J. Arafune, P.G.O. Freund, and C.J. Goebel, J. Math. Phys. {\bf 16}, 433 (1975).

\bibitem[22]{kn:22} S. Coleman, {\it New Phenomena in Subnuclear Physics}, Proc. 1975 Int. School of Physics `Ettore Majorana', ed A Zichichi, New York Plenum, 297 (1975).

\bibitem[23]{kn:23} L.D. Faddeev, {\it Nonlocal, Nonlinear and Nonrenormalisable Field Theories}, Proc. Int. Symp., Alushta, Dubna: Joint Institute for Nuclear Research, 207 (1976); Lett. Math. Phys. {\bf 1}, 289 (1976).

\end{thebibliography}
\end{document}